\documentclass[%
superscriptaddress,
amsmath,amssymb,
aps,
prb,
showpacs,
twocolumn,
floatfix,
longbibliography
]{revtex4-2}

\usepackage{graphicx}% Include figure files
\usepackage{dcolumn}% Align table columns on decimal point
\usepackage{bm}% bold math
\usepackage{amssymb}
\usepackage{amsmath}
\usepackage{commath}
\usepackage{graphicx} %remove demo to get the image
\usepackage[utf8]{inputenc}%latin1
\usepackage{xcolor}
\usepackage[separate-uncertainty = true]{siunitx}
\usepackage{hyperref}
\hypersetup{
    colorlinks=true,
    linkcolor=blue,
    filecolor=blue,
    urlcolor=cyan,
    }

\begin{document}

\preprint{APS/123-QED}

\title{Thermoelectric cooling properties of a quantum Hall Corbino device}% Force line breaks with \\

\author{Juan Herrera Mateos}
\affiliation{International Center for Advanced Studies and ICIFI, ECyT-UNSAM,  25 de Mayo y Francia, 1650 Buenos Aires, Argentina}
\author{Mariano A. Real}
\affiliation{Instituto Nacional de Tecnolog\'{\i}a Industrial, INTI and INCALIN-UNSAM, Av. Gral. Paz 5445, (1650) Buenos Aires, Argentina}
\author{Christian Reichl}
\affiliation{Solid State Physics Laboratory, ETH Z\"urich, CH-8093 Z\"urich, Switzerland}
\author{Alejandra Tonina}
\affiliation{Instituto Nacional de Tecnolog\'{\i}a Industrial, INTI and INCALIN-UNSAM, Av. Gral. Paz 5445, (1650) Buenos Aires, Argentina}
\author{Werner Wegscheider}
\affiliation{Solid State Physics Laboratory, ETH Z\"urich, CH-8093 Z\"urich, Switzerland}
\author{Werner Dietsche}
\affiliation{Solid State Physics Laboratory, ETH Z\"urich, CH-8093 Z\"urich, Switzerland}
\affiliation{Max-Plack-Institut f\"ur Festk\"orperforschung, Heisenbergstrasse 1, D-70569 Stuttgart, Germany}
\author{Liliana Arrachea}
\affiliation{International Center for Advanced Studies and ICIFI, ECyT-UNSAM, 25 de Mayo y Francia, 1650 Buenos Aires, Argentina}

\date{\today}

\begin{abstract}
We analyze the thermoelectric cooling properties of a Corbino device in the quantum Hall regime on the basis of experimental data of electrical conductance. We focus on  the cooling power and the coefficient of performance within and beyond linear response.  Thermovoltage measurements in this device reported in {\em Phys. Rev. Applied, {\bf 14} 034019  (2020)} indicated that the transport
takes place in the diffusive regime, without signatures of effects due to the electron-phonon interaction in a wide range of temperatures and filling factors. In this regime, the heat and charge currents by electrons  can be described by a single transmission function. We infer this function from experimental data of conductance measurements and we calculate the cooling power and the coefficient of performance for a wide range of filling factors and temperatures, as functions of the thermal and electrical biases. We predict an interesting cooling performance in  several  parameter regimes.
\end{abstract}

%Use showkeys class option if keyword
                              %display desired
\maketitle

\section{\label{sec:intro}Introduction}
Thermoelectric phenomena in  quantum mesoscopic devices is a topic of great interest for some years now\cite{Benenti2017Jun,Giazotto2006Mar,ozaeta2014,Marchegiani2020}. The relevant regime takes place at very low temperatures, typically in the subkelvin range. Besides the  interest in  the fundamental physical properties, research in this field is motivated by the emergent development of quantum technologies. For instance, all the proposals for  implementing quantum computation in solid-state devices rely on systems that operate at cryogenic temperatures, being one of such platforms the fractional quantum Hall system\cite{yu,nayak}. One of the challenges in this context is the removal of heat produced in the operation of these devices.

A significant amount of work has been published on thermoelectricity in quantum coherent systems. Paradigmatic examples are quantum dots\cite{Prance2009Apr,Zhang2015May,Entin-Wohlman2015Feb,Roura-Bas2018Nov,Venturelli2013Jun,Schulenborg2017Dec,Sanchez2018Dec,grifoni,misha,aligia}, nanowires\cite{rossella1,rossella2} quantum point contacts\cite{vanHouten1992Mar,Whitney2013Aug,Whitney2015Mar,yamamoto17}  and  topological edge states of the quantum Hall and the quantum spin Hall regimes\cite{Rothe2012,Hwang2014Sep,Sanchez2015Apr,Hofer,Vannucci2015Aug,Vannucci2016,Roura-Bas2018Feb,boehling18,Kheradsoud2019Aug,Sanchez2019Jun,Gresta2019Oct,Blasi2020,hajiloo2020}. In these cases transport takes place through a few ballistic channels. In two-terminal configurations, for very low temperatures, such that electron-phonon interaction does not play a relevant role, charge and thermal currents, as well as the thermoelectric response are fully characterized by a transmission function that describes the transport properties of the electrons through the device \cite{Hicks1993May,Hicks1993Jun,Mahan1996Jul,Heremans2013Jun,Whitney2013Aug,Whitney2015Mar,Benenti2017Jun}. 

\begin{figure}
	\centering
	\includegraphics[width=\columnwidth]{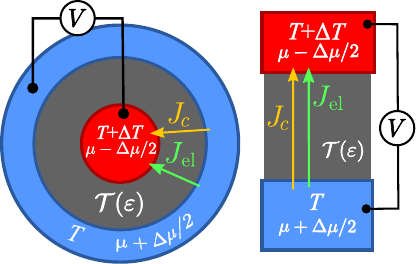}
	\caption{Left panel: sketch of the Corbino ring. The spectrum  of Landau levels define a conductor characterized by a transmission function ${\cal T}(\varepsilon)$ between two biased reservoirs with a temperature difference $\Delta T$. Applying an external voltage $V$ a chemical potential difference  induced, leading to a cooling thermal current $J_{\rm c}$ and an electrical current $J_{\rm el}$.  Right panel: equivalent  two terminal configuration, characterized by the same transmission probability  and identical temperature and voltage biases.}
	\label{fig:fig1}
\end{figure}

Thermoelectricity  in  the quantum Hall regime in the Corbino geometry has been investigated theoretically\cite{barlasyang,morf} and experimentally\cite{vanzalinge-2003,kobayakawa2013diffusion}, as well as in bar configurations\cite{chickering2010,chickering2013}.
In a recent work, experimental evidence of a significant thermoelectric response in a quantum Hall Corbino device was presented \cite{real2020prappl}. The theoretical modeling of this experimental data is consistent with a diffusive mechanism of the electron transport, free from the effects of electron-phonon interaction, accounting for the thermoelectric response within partially  filled Landau levels. This description is based on  a single transmission function, which  determines all the transport coefficients along the radial direction of the Corbino disk,  akin  the case of the ballistic regime. Such a simple and successful interpretation of the experimental data along with the lack of effects originated in the electron-phonon interaction place the Corbino geometry as an excellent candidate for the practical implementation of 
heat--work conversion mechanism and cooling.
In contrast, the bar geometry necessarily deals with the longitudinal and transverse directions, resulting in weaker thermoelectric responses which are 
 significantly more cumbersome from the point of view of the theoretical description\cite{barlasyang,chickering2010,chickering2013}. In addition, effects like the phonon drag, which are less controlled than the transport due to electrons, seem to play a relevant role in quantum Hall bars\cite{fletcher1988,fromhold1993phonon,maximov2004low}.

A necessary ingredient for the thermoelectric  heat-work conversion is the existence of energy filters, implying spectral properties of the device in which the transport of electrons and holes is asymmetric. This implies
a transmission probability  having a profile with rapid changes as a function of the energy. The mathematical properties  that a transmission function must satisfy for an efficient thermoelectric performance  were analyzed  in several works \cite{Mahan1996Jul,Whitney2013Aug,Whitney2015Mar,Samuelsson2017Jun,Gresta2019Oct,hajiloo2020}.

The aim of the present contribution is to present a detailed theoretical study of the cooling properties of the  Corbino device investigated in Ref.~\onlinecite{real2020prappl} on the basis of the transmission function inferred from the experimental data for the conductance recorded in that work. 
A sketch of the considered setup is shown in Fig.~\ref{fig:fig1}. The two-dimensional electron system threaded by an external high magnetic field is constructed with a circular shape, hosting a central hot region, as indicated in the left panel. The system can be  electrically biased by means of a voltage difference. The electrons in the quantum Hall regime are accommodated in Landau levels separated by a gap, which  play the role of energy filters defining the thermoelectric performance. 
The possibility of realizing thermoelectric cooling with the Corbino geometry was previously suggested in Ref.~\cite{Giazotto_2007},  where the term ``Landau cooling" has been coined.
The device effectively behaves as  a two-terminal setup characterized by  a transmission function, as indicated in the right panel of the Figure.  The appealing of using the Landau levels for the
thermoelectric heat--work conversion mechanism is the simplicity of its implementation in comparison to  using edge states, which requires more complex structures with quantum point contacts and quantum dots.

The work is organized as follows. In section II we present the procedure to infer the transmission function from the data of the conductance and we define the theoretical treatment. In section III we analyze the regime for which cooling is possible. The performance is analyzed in Section IV and Section V is devoted to summary and conclusions. 

\section{Theoretical approach}
\subsection{Inference of the transmission function from experimental data}
In Ref.~\cite{real2020prappl} measurements for the conductance and the thermovoltage of a Corbino device with a central heater were reported. The 
experimental data was successfully  explained on the basis of calculations of the Onsager transport coefficients in terms of a transmission function, which was inferred from the conductance.
Here, we will also infer the transmission function following a similar procedure in order to evaluate the cooling power. 
We focus here on the high-magnetic field regime, above \SI{1}{\tesla}, where the contribution of the different Landau levels is clearly resolved in the data of the conductance. In Ref. \onlinecite{real2020prappl}  magnetic fields below \SI{1}{\tesla} --corresponding to filling factors above $\nu=10$ for the measured sample--  were also analyzed. In that regime, the transmission function calculated on the basis of a simple model of Landau levels with a widening due to disorder\cite{jonsonThermo,barlasyang} was found to reproduce accurately the data for the conductance and the thermovoltage. That model failed to accurately reproduce the data for lower filling factors, where the transport coefficients show very peculiar structure, which is very specific of the different filling factors. The particular features in this case also depend on the sample. The aim of the present work is to predict as accurately as possible the cooling properties of the device. Hence, we use an inference procedure to extract the transmission function from the data of the conductance $G_{\rm exp}(B,T_{0})$, as a function of the magnetic field $B$ at a low temperature $T_0$.
We summarize here the main steps followed in Ref.~\cite{real2020prappl} to this end.

\begin{figure}
	\centering
	\includegraphics[width=\columnwidth]{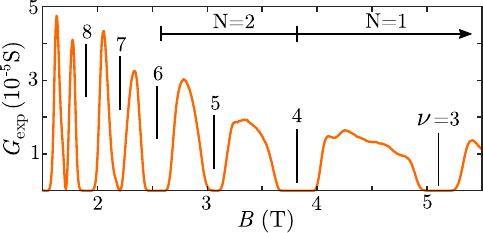}
	\caption {Conductance measurement of a Corbino device as a function of the magnetic field for $T_0=\SI{269}{\milli\kelvin}$. Filling fractions $\nu$ and Landau levels N are indicated. 
	In the text, we will refer to the ``peak $\nu$" for ``the peak before the filling factor $\nu$". The measurement shown corresponds to a Corbino sample of \SI{3.5}{\milli\meter} in diameter, having several concentric ohmic contacts and a central heater. The particular Corbino ring measured in this figure being of \SI{600}{\micro\meter} and \SI{1600}{\micro\meter} inner and outer ring respectively. The sample was grown by molecular beam-epitaxy on a GaAs wafer having a single 2DES. From van-der-Pauw geometry measurements on test pieces a mobility of \SI{21E6}{\square\centi\meter\per\volt\per\second} and an electron density of \SI{2.0E11}{\per\square\centi\meter} was determined at \SI{1.3}{\kelvin} in the dark\cite{real2020prappl}.
	}
	\label{fig:fig2}
\end{figure}

In the setup sketched in the left panel of Fig.~\ref{fig:fig1}, 
the conductance is obtained by measuring the electric current as a function of the magnetic field with the inner and outer reservoirs at the same temperature $T_0$. The system is biased with a small voltage difference $V$ applied radially. The experimental conductance $G_{\rm exp}(B, T_0)$ is directly given by the ratio between this current and $V$. As indicated in the right panel of the Figure, the setup can be effectively regarded as a two-terminal configuration, where the transport properties are described by the transmission probability ${\cal T}(\varepsilon)$. 
Formally, the corresponding conductance at a temperature $T$ is related to this function through 
\begin{equation}\label{cond}
    G(\mu,T)=- \frac{e^2}{h} \int_{-\infty}^{+\infty}  d\varepsilon\, {\cal T}(\varepsilon) \frac{ \partial f(\varepsilon)}{\partial \varepsilon},
\end{equation}
where 
$f(\varepsilon)= 1/(e^{(\varepsilon-\mu)/k_B T}+1)$ 
is the Fermi-Dirac distribution function, which depends on the reference chemical potential $\mu$ and the temperature $T$. 
 Deep in the  quantum Hall regime, where the contributions of the different integer filling fractions are clearly separated one another, the Fermi energy is related to the magnetic field $B_{\nu}=n_e h /e \nu $ corresponding to the filling fraction $\nu$ as $\varepsilon_F=\hbar e B_{\nu}/2 m^*$,  being $n_e$ the electron density of the sample. In a real sample, Landau levels are broadened
and the previous relations extend to a range of magnetic fields around $B_{\nu}$.
On the other hand, for low enough temperatures, the derivative of the Fermi function can be approximated as $- \partial f(\varepsilon)/\partial \varepsilon \rightarrow \delta(\varepsilon -\mu)$. Hence, Eq.~(\ref{cond}) can be expressed as $G(\mu,T)= (e^2/h)\,{\cal T}(\mu)$, while $\mu=\varepsilon_F$. 
Therefore, assuming that the temperature $T_0$ is low enough, 
we can calculate the transmission function from the experimental conductance $G_{\rm exp}(B,T_{0})$  as follows,
\begin{equation} \label{tau-g}
    {\cal T}(\varepsilon) =\frac{h}{e^2} G_{\rm exp}(B,T_{0}),
\end{equation}
with $\varepsilon=\hbar e B/2 m^*$. 
Experimental conductance results at a temperature $T_0=\SI{269}{\milli\kelvin}$ are shown in Fig.~\ref{fig:fig2}.
We can see in the figure a profile of several features separated by an energy gap. Each of these features is associated to a filled Landau level (identified with a label $N$), which
is widened as a consequence of disorder in the sample and split because of the Zeeman interaction.
The corresponding integer filling factors are indicated with $\nu$ in the figure. 
Notice that while this inference of the transmission function is an exact procedure for a conductance recorded at $T_0=0$, there is a distortion as a consequence of the finite temperature at which the experiment is performed, which we will neglect in what follows.
The alignment of the Landau levels to the reservoirs chemical potential is modified by means of the magnetic field. Hence, it can be regarded as a knob to tune the transport properties of the device.  

This profile contains features similar to smoothed step functions separated by wells and also Lorentzian-type peaks. Since the cooling properties of  transmission functions with these characteristics were recently carefully analyzed in Ref.~\cite{hajiloo2020}, we will take into account several insights from that work as a reference for our study. 

\subsection{Cooling current and electrical power}\label{sec:currents}
We now consider the situation where, in addition to a voltage bias $V$, a temperature bias associated to a thermal load,   is radially  applied. In order to focus on a configuration close to the experimental one studied in Ref.~\cite{real2020prappl}, which contained a heater in the center of the sample, we present in Fig.~\ref{fig:fig1} the hottest reservoir at the center. 
 However, we stress that the forthcoming analysis does not depend on whether the hot reservoir is placed in the central region of the Corbino or in the external rim. Furthermore, we highlight the fact that the present configuration can be actually regarded as a two-terminal setup under the effect of a thermal bias $\Delta T$, and an electrical bias $\Delta \mu= eV$, as indicated 
in the right-hand side of Fig.~\ref{fig:fig1}. In this representation it is clear that the position of the cold and hot reservoirs can be exchanged. 

Following Ref.~\cite{hajiloo2020}, we focus on the heat current transported by the electrons, exiting the reservoir at the lowest temperature $T$. In terms of the transmission function, it reads
\begin{equation}\label{heat}
    J_{\rm q}=  - \frac{1}{h} \int_{-\infty}^{+\infty}  d\varepsilon {\cal T}(\varepsilon) \left( \varepsilon - \mu - \frac{\Delta \mu}{2} \right) 
    \Delta f(\varepsilon),
\end{equation}
being $\Delta f(\varepsilon)= f_{\rm c}(\varepsilon)- f_{\rm h}(\varepsilon) $, the difference between the Fermi functions corresponding to the cold/hot (c/h) reservoir at temperatures
$T_{\rm c} = T$, $T_{\rm h} = T+\Delta T$ and  chemical potentials $\mu_{\rm c}=\mu + \Delta\mu/2$ and 
$\mu_{\rm h} = \mu -\Delta\mu/2$, respectively. The heat flux given by  Eq. (\ref{heat}) is directly proportional to the entropy flux entering or exiting the coldest reservoir due to the particle flux through the device. In the absence of inelastic scattering processes, the latter is fully characterized by its transmission  probability
${\cal T}(\varepsilon)$. 
The two sources of inelastic scattering in these systems are electron-phonon and electron-electron interactions. In Ref. \onlinecite{real2020prappl}, no signatures of electron-phonon interactions in the thermoelectric response up to temperatures as high as \SI{1.5}{\kelvin} where found. Electron-phonon interaction typically generates the mechanism of phonon-drag, which has been observed previously in the bar geometry \cite{fletcher1988,fromhold1993phonon,maximov2004low}, but seems to be absent or very weak in the Corbino geometry \cite{kobayakawa2013diffusion, real2020prappl}. Electron-electron interaction for large voltage and temperature biases
can lead to non-equilibrium corrections to the transmission function ${\cal T}(\varepsilon)$ calculated from Eq.~(\ref{tau-g}). These corrections are expected to be important within the non-linear regime in situations where the screening is poor \cite{screen1,screen2,screen3,screen4,screen5}. We will neglect these effects here, since we focus on the transport through partially filled Landau levels, in which case the screening is expected to be efficient. Furthermore, we anticipate that the relevant operation regime is at low electrical and thermal biases, in which case non-linear effects are not dominant.

The heat flux described by Eq. (\ref{heat}) corresponds to a cooling process when it exits the coldest reservoir. 
We define the {\em cooling power} as
\begin{equation}\label{jc}
    J_{\rm c} \equiv J_{\rm q}, ~{\rm if} \; J_{\rm q} \geq 0, ~~~~~~~~~ J_{\rm c} \equiv 0, ~
    {\rm otherwise}.
\end{equation}
Similarly to other works, we will consider as a reference the Pendry quantum bound\cite{Whitney2013Aug, Whitney2015Mar, Pendry1983Jul} 
\begin{equation}
    J_{\rm qb}= \frac{\pi^2 k_B^2}{6h} T^2,
 \label{Pendry}
\end{equation}
which defines the maximum heat current that can be transported through a quantum channel of unitary transmission. 

\begin{figure}
	\centering
	\includegraphics[width=\columnwidth]{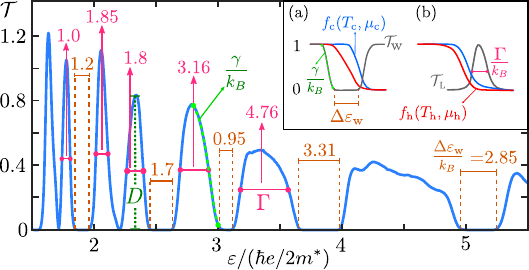}
	\caption{
	Transmission function inferred from the measured conductance shown in Fig.~\ref{fig:fig2}. 
	Equivalent temperature estimates in kelvin for the gaps (orange), half height width of the Landau levels $\Gamma/k_B$ (pink), smoothing parameter $\gamma /k_B$ and associated height parameter $D$ are indicated. These will impact the cooling range as well as the the optimal operational conditions (see text). \\
	\textit{Inset:} (a) Well-like transmission function ${\cal T}_{\rm w}$ (grey) including the  Fermi functions $f_{\rm h}$ for the hot (red) and the cold  $f_{\rm c}$ (blue) reservoirs. The width of the well is $\varepsilon_{\rm w}$. (b) Lorentzian-like transmission function ${\cal T}_{\rm L}$ of width $\Gamma$ and height $D=1$.
	%Some approximate values of $\gamma/4k_B$ for filling factors 8 to 4 are: \SIlist{190;330;360;600;840}{\milli\kelvin} respectively.
	}
	\label{fig:fig3}
\end{figure}

The thermal and electrical biases induce an electron flux $J_{\rm el}$, and  have an associated electrical power $P$, which are given by 
\begin{equation}\label{pch}
     J_{\rm el} = - \frac{1}{h} \int_{-\infty}^{+\infty}  d\varepsilon \, {\cal T}(\varepsilon)
    \Delta f(\varepsilon), ~~~~~~P= \Delta\mu J_{\rm el}.
\end{equation}
We stress that Eqs. (\ref{jc}) and (\ref{pch}) do not assume linear response. Hence, we will analyze the response for a wide range of electrical and thermal biases. We will also include a discussion on the performance within the linear-response regime for comparison.

\subsection{Cooling performance}
It is natural to quantify the efficiency of the cooling operation in terms of the coefficient of performance, defined as
\begin{equation}
    {\rm COP}=\frac{J_{\rm c}}{P},
\end{equation}
with the definitions of the cooling power and the invested electrical power given by Eqs.  (\ref{jc}) and (\ref{pch}), respectively. This quantity is bounded by the 
Carnot limit $\eta_C=T/\Delta T$. 

\subsection{Linear-response regime}
A particular important regime corresponds to small amplitudes of the thermal and electrical biases, where $\Delta T/T$ and $\Delta\mu/k_B T$ are small enough to  justify an expansion up to linear order in these quantities in the expressions of the heat and charge currents
defined, respectively, in Eqs. (\ref{heat}) and (\ref{pch}). The result is
\begin{equation}\label{thermo}
\left(
\begin{array}{c}
J_{\rm el}/e \\
J_{\rm q} 
\end{array}
\right)  =  \left(
\begin{array}{cc}
{\cal L}_{11} &{\cal  L}_{12}  \\
{\cal L}_{21} &{\cal L}_{22} 
\end{array}
\right) 
\left(
\begin{array}{c}
\Delta\mu /k_B T\\
\Delta T/k_B T^2
\end{array}
\right).
\end{equation}
where $\hat{\cal L}$ is  the Onsager matrix, with elements calculated from the transmission function as follows
\begin{equation}\label{onsa}
{\cal L}_{ ij}= - T \int \frac{d \varepsilon}{h} \frac{\partial f (\varepsilon)}{\partial \varepsilon} \left(\varepsilon-\mu \right)^{i+j-2} {\cal T}(\varepsilon).
\end{equation}
Notice that ${\cal L}_{12}={\cal L}_{21}$, since 
the present device effectively behaves like a usual two-terminal one, as sketched in Fig.~\ref{fig:fig1}.

In this regime, the quantity characterizing the maximum COP for a fixed value of $\Delta T$ is parametrized by the figure of merit $ZT={\cal L}_{21}^2/\mbox{Det}{\hat{\cal L}}$ through\cite{Benenti2017Jun} 
\begin{equation}\label{eq:ZTaprox}
{\rm COP}=\eta_C \frac{\sqrt{ZT+1}-1}{\sqrt{ZT+1+1}}.
\end{equation}
  Hence, the ideal Carnot limit is achieved for $ZT \rightarrow \infty$. 

The usual situation in thermoelectricity is that devices for which the COP is high, the cooling power is low and viceversa \cite{Mahan1996Jul,Benenti2017Jun,Whitney2013Aug,Whitney2015Mar,Samuelsson2017Jun,Gresta2019Oct,hajiloo2020}. Therefore, we will analyze separately the following aspects: (i) under which conditions is cooling possible in the device, (ii) the conditions for which a large cooling power is expected and (iii) the conditions for a large COP.

\subsection{Conditions for cooling}\label{sec:cool}
Before going to the concrete analysis of the cooling properties of the sample for which the conductance is reported in Fig. \ref{fig:fig2}, it is useful to summarize the outcome of Ref. \cite{hajiloo2020} regarding the properties of ${\cal T}(\varepsilon)$ leading to cooling in transmission functions containing peaks, steps and well-type features. 

The optimal profile leading to  the highest cooling power is a single rigid step function of the form ${\cal T}_{\rm step} (\varepsilon) = D \theta (\varepsilon - \varepsilon_{\rm w})$. The optimal operational condition corresponds to the chemical potential of the coldest reservoir coinciding with the onset of the step, $\mu_{\rm c}= \varepsilon_{\rm w}$, and large electrical bias, with $\mu_{\rm h}$ as far apart as possible from the step. This configuration of chemical potentials enables the thermoelectric compensation of the effect of the heat flow because of the temperature bias,  leading to the maximum bound for the cooling power, $J_{\rm c}^{\rm max} = D J_{\rm qb}/2$. 

If, instead of a sharp step, the transmission function has the shape characterized by a smoothness parameter $\gamma$, such that the $\theta$-function in ${\cal T}_{\rm step}(\varepsilon) $ is substituted by a smoothed step-function
$\Theta_{\gamma}(\varepsilon)=1/(1+e^{-\varepsilon/\gamma})$,
the maximum achievable cooling power decreases in reference to $J_{\rm c}^{\rm max}$ by a factor
 $D J_{{\rm qb}}(\gamma/k_B T)^2/4$, which  can be achieved for $T>\gamma/4 k_B $ and similar configurations of the chemical potentials as in the case of a sharp step. Furthermore, for temperatures $T<\gamma/4 k_B $, cooling is strongly suppressed and it is possible only for small $\Delta \mu$.
  %\textcolor{blue}{Note that this is the case for even filling factors, while for odd ones there is a competition between this value and $\Delta\varepsilon_w/k_B$.} The cooling power is even further  reduced for $T < \gamma/4 k_B $ 
 %(the thermal width of the transport window is $\sim 4 k_B T$). 
 
 On the other hand,  the cooling operation becomes strongly limited at high temperatures by the existence of well-type features. A sketch of such transmission function, denoted by ${\cal T}_{\rm w}(\varepsilon)$
 and characterized by a typical energy $\Delta \varepsilon_{\rm w}$, is indicated in the inset (1) of Fig.~\ref{fig:fig3}, along with the two Fermi functions $f_{\rm c}(\varepsilon)$ and $f_{\rm h}(\varepsilon)$. In such cases, starting from a situation where the chemical potential $\mu_{\rm c}$ is fixed at one of  the step functions, the amplitude of the 
  bias voltage --applied to compensate the 
 thermal bias-- is limited by the existence of the mirror step function. Hence, the cooling operation in this case is limited by $T_{\rm h},~ T_{\rm c} \sim \Delta \varepsilon_{\rm w}/k_B $. The inset of Fig.~\ref{fig:fig3} shows a configuration with the chemical potential of the cold reservoir selected at the step with positive slope, in which case $\Delta\mu=\mu_{\rm c}-\mu_{\rm h} >0$. An identical result can be obtained when $\mu_{\rm c}$ is
 placed at the step with negative slope (the mirror-related step) and $\Delta\mu= \mu_{\rm c}-\mu_{\rm h} < 0$. 
 
 In the case of sharp peaks (Lorentzian-like) features, there are other limiting properties in the operation. An example of such a transmission function is sketched in the inset (b) of Fig. \ref{fig:fig3}, denoted by ${\cal T}_{\rm L}(\varepsilon)$ and characterized by a width $\Gamma$. In the present case, cooling can also be achieved for $\mu_{\rm c}$ placed at the
 positive slope of the peak and $\Delta\mu= \mu_{\rm c}-\mu_{\rm h}>0$ applied to compensate the thermal bias or, equivalently, when $\mu_{\rm c}$ is placed at the
 negative slope of the peak and $\Delta\mu =\mu_{\rm c}-\mu_{\rm h} <0$ applied. The 
cooling operation is limited in this case  to the regime where  $T_{\rm c}$,~$T_{\rm h} < \Gamma / 4 k_B $ and is possible only for small values of $|\Delta \mu|$.

\begin{figure}
	\centering
	\includegraphics[width=\columnwidth]{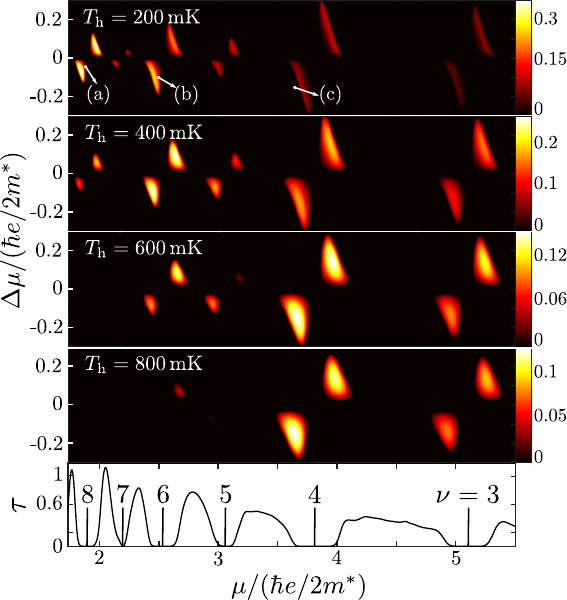}
	\caption{Contour plots for the cooling power normalized by the quantum bound, $J_{\rm c}/J_{\rm qb}$ as a function of $\mu/(\hbar e/2m^*)$ and $\Delta\mu/(\hbar e/2m^*)$ at a fixed relation $\Delta T/T_{\rm h}=0.1$. ${\cal T}(\varepsilon)$ within the same range of energies is shown as a reference in the bottom panel. Note that the color scale changes on each panel.
	We indicate in the top panel three points in $\mu$ and $\Delta\mu$ space that will be used over the text, having values  (in units of $\hbar e / 2m^*$) of $\mu_{(a)}=1.85$, $\Delta\mu_{(a)}=-0.064$; $\mu_{(b)}=2.459$, $\Delta\mu_{(b)}=-0.095$ and $\mu_{(c)}=3.654$, $\Delta\mu_{(c)}=-0.144$.
	} 
	\label{fig:range1}
\end{figure}

\section{Cooling power}
In Fig.~\ref{fig:fig3}, the parameters characterizing the different features of the transmission function ${\cal T}(\varepsilon)$ are indicated. In particular, 
widths of the main peaks and wells, respectively $\Gamma$ and $\Delta \varepsilon_{\rm w}$ are indicated in units of temperature in order to facilitate the comparison  of these properties with the thermal width entering the function $\Delta f$ in Eq. (\ref{jc}), which we consider to be  $\sim 4 k_B T$. The height $D$ of each feature can be simply read in the vertical axis.

At high temperatures, the main limitation for the cooling range is the width of gaps between the different Landau levels. If the 
gap between two consecutive Landau levels is
of the same order of magnitude of the transport window, cooling is no longer possible.   

\begin{figure}
	\centering
	\includegraphics[width=\columnwidth]{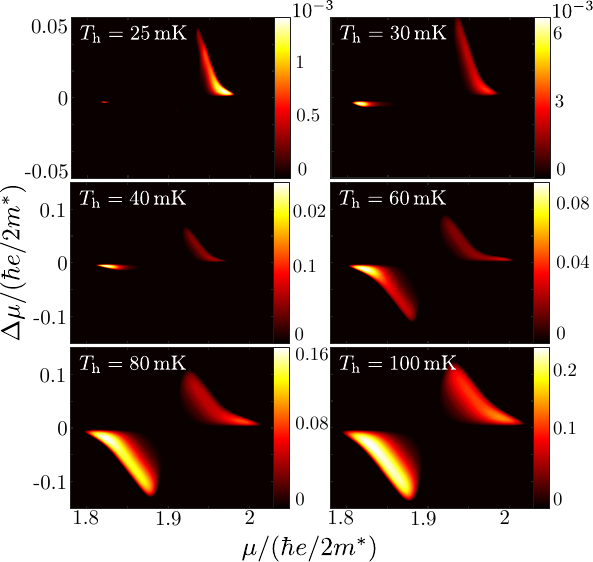}
	\caption{Similar to Fig.~\ref{fig:range1} with focus on the low temperature regime. Contour plots corresponding to the vicinity of $\nu = 8$. }
	\label{fig:range2}
\end{figure}
The  cooling power normalized by the quantum bound, $J_{\rm c}/J_{\rm qb}$,
as a function of $\mu$ and $\Delta \mu$
% =\mu_{\rm c}- \mu_{\rm h}= eV
in a range of temperatures 
$\SI{200}{\milli\kelvin} < T < \SI{800}{\milli\kelvin}$ is shown in Fig.~\ref{fig:range1} for a fixed ratio $\Delta T/T_{\rm h} = 0.1$. In this Fig. we can identify different regions of values of chemical potentials and voltages for which cooling is possible in the device within this temperature regime. We see that they are localized close to values of energies for which ${\cal T}(\varepsilon)$ has rapid changes and the profile of ${\cal T}(\varepsilon)$ is similar to well-type  ${\cal T}_{\rm w}$ and peak-like ${\cal T}_{\rm L}$ transmission functions discussed before. This takes place close to the onset and closing of the different Landau levels. As mentioned in Section \ref{sec:cool},  the sign of $\Delta\mu$ leading to cooling depends on the sign of the corresponding slope of 
${\cal T}(\mu_{\rm c})$. For the lowest temperature, shown in the upper panel, we see that the highest intensity corresponds to the peaks $\nu=6,7$ and 8.
This is because in this temperature range, $T_{\rm c}$ and $T_{\rm h}$ are lower than the width of these peaks, as well as lower than the gaps between the peaks $\nu = 7$ and 8 and between $\nu=5$ and 6.
Under these conditions, the features favoring the strength of the power are the slope ($1/\gamma$) and the height ($D$) of ${\cal T}(\varepsilon)$, and these two peaks are the optimal ones regarding these properties. The estimates are $\gamma/4k_B \sim $ \SIlist{360;329;190}{\milli\kelvin} for $\nu=6,7,8$,  respectively, while $D\sim 1.2$. 
On the other hand, as the temperature grows (see top to bottom panels in the Fig.), the cooling lobes centered on these filling factors become fainter and disappear when the temperature  becomes comparable to the width of the gap between neighboring peaks. In the high-temperature range, the optimal cooling region is shifted towards lower filling factors ($\nu= 3, 4$). The cooling power is decreased in comparison to the previous cases since the slopes and heights are smaller in these cases. However, the range of $\mu, \Delta\mu$ for which cooling is possible is larger for these lower filling factors, since the widths $\Gamma$ of the Landau levels and the gaps $\Delta \varepsilon_{\rm w}$ become larger.

A similar analysis with focus on the low-temperature regime is presented in Fig.~\ref{fig:range2}. These plots show the cooling power within a range of temperature
$\SI{25}{\milli\kelvin}<T<\SI{100}{\milli\kelvin}$. The optimal range of magnetic fields (or equivalent chemical potentials) in this case corresponds to the one indicated with (a) in Fig.~\ref{fig:range1} and is centered between $\nu=7$ and 8. These peaks of ${\cal T}(\varepsilon)$ have  approximately the same width $\Gamma/4 k_B \sim \SI{460}{\milli\kelvin}$ and a corresponding slope $\gamma/4 k_B \sim \SI{360}{\milli\kelvin}$. 
The range of temperatures covered in Fig.~\ref{fig:range2} have an associated thermal broadening, $\sim 4 k_B T_{\rm c}$,
%$\sim 4 k_B T_{\rm c}$
 which is smaller than $\gamma$. 
This explains the low values of the cooling power, which shows a decreasing behavior as the temperature decreases. Notice that the highest values of $J_{\rm c}$ are  achieved in this region of chemical potentials close to $T_{\rm h} \sim \SI{200}{\milli\kelvin}$ (shown in the top panel of Fig.~\ref{fig:range1}), which corresponds precisely to the regime where $T_{\rm c}$ becomes larger than $\gamma /4 k_B$.

\begin{figure}
	\centering
	\includegraphics[width=\columnwidth]{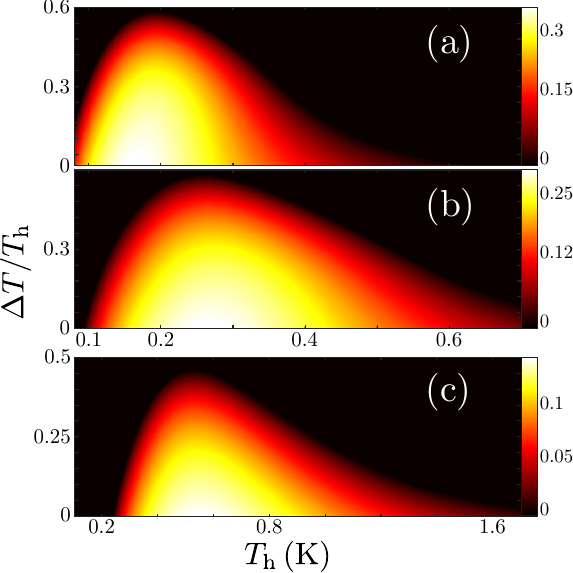}
	\caption{
	Contour plots for the cooling power normalized by the quantum bound. The fixed values of $\mu$ and $\Delta\mu$ are marked in Fig.~\ref{fig:range1} and correspond (in units of $\hbar e / 2m^*$) to $\mu_{(a)}=1.85$, $\Delta\mu_{(a)}=-0.064$; $\mu_{(b)}=2.459$, $\Delta\mu_{(b)}=-0.095$ and $\mu_{(c)}=3.654$, $\Delta\mu_{(c)}=-0.144$.\\
	}
	\label{fig:ContornosTemp}
\end{figure}
We close the analysis of the cooling power  with the results shown in Fig.~\ref{fig:ContornosTemp}. These correspond to the cooling power as a function of the temperature of the hot bath
$T_{\rm h}$ and the relative temperature difference $\Delta T/T_{\rm h}$ for fixed vales of $\mu$ and $\Delta\mu$. These were chosen to coincide with the ones identified with (a), (b) and (c) in Fig.~\ref{fig:range1} and are the ones leading to the highest cooling power close to $\nu=8,6,4$, respectively. 
From the information of this Fig. we see that cooling at low temperatures is possible only near $\nu=8$, corresponding to panel (a). 
This is precisely the case analyzed in Fig.~\ref{fig:range2}. Similar values of the cooling power can be achieved for the parameters analyzed in panel (b). In this case, cooling is only possible above $T \sim \SI{100}{\milli\kelvin}$. Hence, the low temperature regime analyzed in Fig.~\ref{fig:range2} cannot be reached by cooling with the peak $\nu=6$. The reason is that the transmission function peak in the latter case is smoother than in the case (a) (compare $\gamma/4k_B \sim \SI{625}{\milli\kelvin}$ to $\gamma/4 k_B = \SI{190}{\milli\kelvin}$ of the previous case). 
On the other hand, cooling is possible up to higher temperatures in case (b), up to $T_{\rm h} \sim \SI{750}{\milli\kelvin}$. Since $\Gamma/4 k_B$ as well as $\Delta \varepsilon_{\rm w}$ are larger in (b) than in (a). Similar observations can be done on the cooling power of the Landau level peak below $\nu=4$, which is shown in panel (c). In that case, the range of temperature for cooling  is the largest one, as a consequence of the large values of $\Gamma$ and $\Delta \varepsilon_{\rm w}$ associated to this Landau level. The onset for cooling is at a higher temperature than in the previous cases, $T_{\rm h}\sim \SI{200}{\milli\kelvin}$ as a consequence of the larger smoothness $\gamma /4 k_B \sim \SI{720}{\milli\kelvin}$. The largest achieved values for the cooling power are a bit lower than in the previous cases, as a consequence of the smaller height $D$ of ${\cal T}(\varepsilon)$.

The absolute maximums of the cooling power as a function of temperature are determined by the smoothness parameter $\gamma$ of the different Landau levels
and the width $\varepsilon_{\rm w}$ of the gap. Therefore,  they take place  at $\nu = 8$ for temperatures below \SI{300}{\milli\kelvin},  at $\nu = 6$ and  at $\nu = 4$ for temperatures above $\sim \SI{600}{\milli\kelvin}$. It is also interesting to note that the specific values o $\mu$ and $\Delta\mu$  do not change considerably and the temperature changes. For example if we focus on the lobe (a) in Fig.~\ref{fig:range1} the cooling area  decreases as the temperature increases. However, the values of $\mu,~\Delta \mu$ corresponding to the maximum position remain almost the same until the temperature is high enough that the maximum cooling power takes place at the  filling factor, indicated with (b). This happens at a temperature, $\sim \SI{400}{\milli\kelvin}$.
%This means that during an experiment one could select a specific magnetic field and bias voltage to be near a maximum power and will remain almost ideal for the temperature range for that cooling lobe.
For even higher temperatures $\sim\SI{600}{\milli\kelvin}$, a similar effect takes place between the lobes associated to (b) and (c).

\section{Cooling efficiency}
\begin{figure}
	\centering
	\includegraphics[width=\columnwidth]{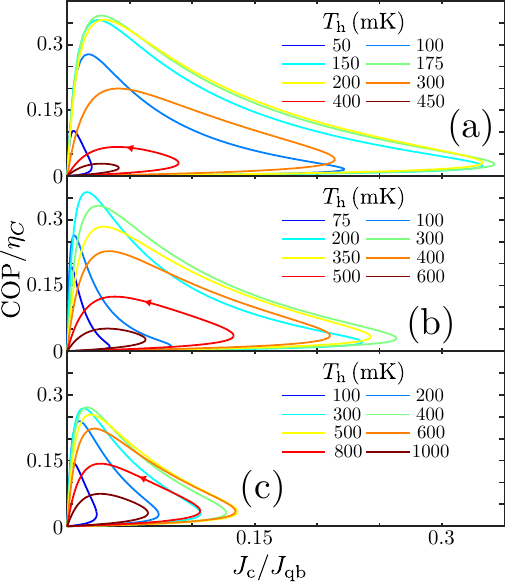}
	\caption{Coefficient of performance normalized by the Carnot efficiency ($\eta_C=T_{\rm c}/\Delta T$) versus the cooling power normalized by the quantum bound ($J_{\rm qb}$), for varying $\Delta\mu/(\hbar e / 2 m^*)$ from  $-0.25$ to $0.25$. The values of $\mu/(\hbar e / 2 m^*)$ are 1.844,  2.459 and 3.654, corresponding to points (a), (b) and (c) indicated in Fig.~\ref{fig:range1}, respectively.
	$\Delta T/T_{\rm h}=0.1$ in all cases.
}
	\label{fig:COPvsJ}
\end{figure}

The usual situation in  thermoelectric devices is that
the highest efficiencies are achieved for parameters where the cooling power is very low and vice-versa and this is also the case in the present device.  For this reason, it is very useful to represent both quantities at the same time in the form of the so called ``lasso" plots, where COP vs $J_{\rm c}$ are represented for a set of parameters.

Fig.~\ref{fig:COPvsJ} shows ``lasso" representations of  COP vs $J_{\rm c}$  as $\Delta\mu / (\hbar e / 2 m^*)$ changes within a specific interval while $\mu$ correspond to the (a), (b) and (c) positions indicated in Fig.~\ref{fig:range1} and $\Delta T/T_{\rm h}=0.1$ in all cases. As mentioned before, the largest values of  $J_{\rm c}$ correspond to the smallest values of COP and vice-versa, but we can see wide ranges of parameters where both qualifiers achieve interesting values.

So far, we have not introduced any assumption regarding the amplitudes of the thermal and electrical biases. For the case where $\Delta T/T_{\rm c}$ and $\Delta\mu/k_B T_{\rm c}$ are small enough to  justify an expansion up to linear order as in Eq. (\ref{thermo}) the COP is given by Eq. (\ref{eq:ZTaprox}).
\begin{figure}
	\centering
	\includegraphics[width=\columnwidth]{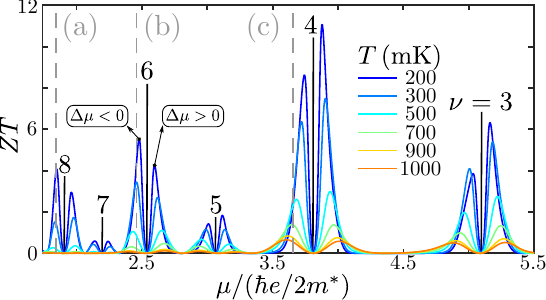}
	\caption{Figure of merit, $ZT$ in the linear-response regime intercepted by the $ \mu/(\hbar e/2m^*)$ values of points (a), (b) and (c) as indicated in Fig. \ref{fig:range1}. We note that the highest performance is in the vicinity of the fill factor $\nu=4$. }
	\label{fig:ZT}
\end{figure}

 The corresponding results are shown in Fig. \ref{fig:ZT}. In order to get cooling, it is necessary to apply a voltage bias with a positive or negative sign,  depending
 on the properties of ${\cal T}(\varepsilon)$, which define the behavior of ${\cal L}_{ij}$. Crucially, the sign of the off-diagonal coefficients, ${\cal L}_{12}={\cal L}_{21}$ depend on the sign of the slope of ${\cal T}(\mu_{\rm c})$. Thus, $\Delta\mu >0$ ($\Delta\mu <0$) for positive (negative) slope of ${\cal T}(\varepsilon)$. The different signs are indicated in Fig. \ref{fig:ZT}: the left side of the peaks correspond to $\Delta\mu<0$ and the right ones to $\Delta\mu>0$. As already mentioned in Ref.~\onlinecite{real2020prappl}, these values of $ZT$ are remarkably high.
  
 Note that the values of $\mu / (\hbar e / 2m ^ *) $ associated to (a), (b) and (c) do not correspond to maximum values of 
 $ZT$, while they are perfectly consistent with the results shown in Fig.~\ref{fig:COPvsJ}. Interestingly, the values of the COP calculated from $ZT$ of Fig.~\ref{eq:ZTaprox} are the same as the maximum ones in Fig.~\ref{fig:COPvsJ} for  $\Delta T/T_{\rm h}=0.1$. This means that the results  of Fig.~\ref{fig:COPvsJ} are very close to the conditions for linear response operation.

\section{Summary and conclusions}
We have used experimental data of the conductance of the Corbino device studied in Ref. \onlinecite{real2020prappl}  to infer the transmission function. In that reference, it was experimentally verified that the thermoelectric transport takes place in the diffusive regime, without effects introduced by the electron-phonon coupling, implying that it can be accurately described in terms of the transmission function so calculated. Here, we have used that transmission function  to evaluate the cooling power and the COP for cooling. 

We have investigated the cooling properties for a wide range of electrical and thermal biases, within and beyond the linear-response regime.
According to our analysis, in the present sample, the cooling power relative to $J_{\rm qb}$ for this system is maximum in the vicinity of filling factor $\nu = 8$,  and the resulting value is: $\left ( J_{\rm c}/J_{\rm qb} \right )^{max} =0.346$
at $\mu/(\hbar e/2m^*)=1.85$,  $T_{\rm h}=170\, {\rm mK}$, for a thermal load $\Delta T= \SI{5.5}{\milli\kelvin}$.
 The chemical potential difference associated to this case results in a bias voltage of $V= \SI{-60}{\micro\volt}$. 
 We found  that the temperature difference can be increased without a significant decrease in the cooling power. 
 
 We have also analyzed the cooling coefficient of performance, as well as the figure of merit characterizing the performance in linear response and we found interesting results, corresponding
 to large COP ($\sim 0.3 \eta_C$ and larger) for several filling factors, mainly $\nu=4,6,8$ in the subkelvin regime at a sizable cooling power. 
 
 Our estimates, did not include the phononic contribution of the thermal conductance. According to Ref.~\onlinecite{chickering2010},
 the latter follows a law $\kappa_{\rm ph}= \alpha_{\rm ph} T^3$. We expect the thermoelectric cooling operation to be particularly useful at low temperatures  when the electronic cooling power 
 $J_{\rm c} \propto T^2$ should overcome the
 phononic one, induced by the thermal bias.

 The specific values obtained from our analysis are associated to the specific sample and device of Ref.~\onlinecite{real2020prappl}. However, we have verified that the orders of magnitude of the different quantities are  valid for a wide range of samples
with similar  mobility and devices with similar dimensions, including samples doped with Cr, in which case the phonon conductance is expected to be reduced \cite{chaudhuri1973thermal}. Therefore, the present results constitute a proof of principle for the applicability of Landau cooling with Corbino devices in cryogenic conditions. On the other hand, once the transmission function is inferred, the roadmap of our analysis, based on the results presented in Ref.~\onlinecite{hajiloo2020}, can be also implemented in any other similar particular realization.

\section{Acknowledgements}

We thank Klaus von Klitzing for his constant interest and support and Janine Splettstoesser for carefully reading the manuscript and constructive comments.

We acknowledge support from INTI (MR, AT) and CONICET (JHM, LA), Argentina. We are sponsored by PIP-RD 20141216-4905 of CONICET, PICT-2017- 2726 and PICT-2018 from Argentina, Swiss National Foundation (Schweizerischer National fonds) NCCR ``Quantum Science and Technology", as well as the Alexander von Humboldt Foundation, Germany (LA).

\bibliography{bibliografia}% Produces the bibliography via BibTeX.

\end{document}